\documentclass[intlimits,twoside,a4paper]{article}

\usepackage[T2A]{fontenc} 
\usepackage[utf8]{inputenc} 

\usepackage[eqsecnum]{cmpj3}

\definecolor{darkblue}{rgb}{0.0, 0.0, 0.55}
\definecolor{darkgreen}{rgb}{0.0, 0.2, 0.13}
\definecolor{darkred}{rgb}{0.55, 0.0, 0.0}

\articletype{A part of the Special collection to the 100th anniversary of the birth of Ihor Yukhnovskii}

\issue{2025}{28}{4}{43502}
\doinumber{10.5488/CMP.28.43502}

\title{Three-dimensional unfrustrated and frustrated quantum Heisenberg magnets. Specific heat study} 

\author[T. Krokhmalskii, T. Hutak, O. Derzhko]{T. Krokhmalskii\orcid{0000-0003-4186-5599}\refaddr{label1}, T. Hutak\orcid{0000-0002-3601-0601}\refaddr{label1}, O. Derzhko\orcid{0000-0002-4187-0518}\refaddr{label1,label2}} 
\addresses{
	\addr{label1}Yukhnovskii Institute for Condensed Matter Physics 
	of the National Academy of Sciences of Ukraine,\\
	Svientsitskii Street 1, 79011 Lviv, Ukraine
	\addr{label2}	Professor Ivan Vakarchuk Department for Theoretical Physics,
	Ivan Franko National University of Lviv,\\ 
	Drahomanov Street 12, 79005 Lviv, Ukraine}


\date{Received July 20, 2025, in final form September 30, 2025}

\begin{document}
	
	\maketitle

\begin{abstract}
We examine the $S=1/2$ Heisenberg magnet on four three-dimensional lattices ---  simple-cubic, diamond, pyrochlore, and hyperkagome ones --- for ferromagnetic and antiferromagnetic signs of the exchange interaction in order to illustrate the effect of lattice geometry on the finite-temperature thermodynamic properties with a focus on the specific heat $c(T)$. To this end, we use quantum Monte Carlo simulations or  high-temperature expansion series complemented with the entropy method. We also discuss a recent proposal about hidden energy scale in geometrically frustrated magnets.
\printkeywords
\end{abstract}



Its our honor and pleasure to contribute to the Condensed Matter Physics journal project celebrating I.~R.~Yukhnovskii's 100th birthday. Two of us (T.~K. and O.~D.) were strongly influenced by his personality since 1980s. We keep in mind so many lovely memories of him, and not only those related to physics but also to everyday life in general. Looking back over those years, we realize how green and simple-minded we were and how good were his pieces of advise. With deep gratitude we acknowledge the kindness that always irradiated from him. 

\section{Introduction}
\label{s1}

Statistical mechanics and thermodynamics are of great importance for understanding the basic nature of the world. Thus, the quantum era was opened by solving four problems, two of which --- the black-body radiation (M.~Planck) and the specific heat of solids (A.~Einstein) --- concern statistical mechanics and thermodynamics \cite{Born2013,Vakarchuk2012,Duncan2019}. As a matter of fact, understanding a specific heat, i.e., the amount of heat required to change the temperature of an electromagnetic field or a material in equilibrium with a bath by unit, broke the boundaries of traditional physics. 

This paper deals with thermal properties of the quantum spin $S=1/2$ Heisenberg model on a lattice. The Hamiltonian of the model reads:
\begin{equation}
\label{01}
H=J\sum_{\langle mn\rangle}{\bf{S}}_{m}\cdot{\bf{S}}_{n}\,.
\end{equation}
Here, $J=1$ corresponds to the antiferromagnetic exchange interaction whereas $J=-1$ corresponds to the ferromagnetic exchange interaction and $\langle mn\rangle$ denotes the neighboring lattice sites $m$ and $n$. Furthermore, spin-1/2 operators ${\bf{S}}_{m}=(S_m^{x},S_m^{y},S_m^{z})$ obey the following commutation relations:
\begin{eqnarray}
\left[S_m^{x},S_n^{y}\right]={\rm i}\delta_{mn}S_m^{z}
\end{eqnarray}
($\hbar=1$)
and two more similar relations with the cyclically shifted order of $x$, $y$, $z$. Obviously, Hamiltonian~(\ref{01}) is a sum of noncommuting terms and therefore  represents an  essentially quantum system.

As usual, to find the (equilibrium) thermodynamic quantities, one has to calculate the partition function
\begin{eqnarray}
\label{03}
Z(T,N)=\sum_{j=1}^{2^N}\exp\left(-\frac{E_j}{T}\right)
\end{eqnarray}
($k_{\rm B}=1$),
where $H\psi_j=E_j\psi_j$, i.e., $\psi_j$ is an eigenstate of the Hamiltonian $H$ with the energy $E_j$. Thermodynamic quantities follow from the Helmholtz free energy per lattice site $f$, which is given by
\begin{eqnarray}
\label{04}
f(T)=\lim_{N\to\infty}\frac{F(T,N)}{N},
\quad
F(T,N)=-T\ln Z(T,N).
\end{eqnarray}
In particular, for the entropy per site $s$ and the specific heat per site $c$ we have:
\begin{eqnarray}
\label{05}
s(T)=-\frac{\partial f}{\partial T},
\quad
c(T)=T\frac{\partial s(T)}{\partial T}.
\end{eqnarray} 

Since the set of eigenvalues $\{E_j/J\}$ is identical for ferromagnetic and antiferromagnetic cases, e.g., for $N=2$ we have a singlet with $E/J=-3/4$ and a triplet with $E/J=1/4$, the partition function (\ref{03}) for $J=1$ and $T>0$ equals to the partition function (\ref{03}) for $J=-1$ and $T<0$. However, it is not clear how this relation can be used in practice. In this study, we consider both signs of the exchange coupling, $J>0$ and $J<0$.

It may be instructive to bear in mind a simpler case of the Ising interaction in equation~(\ref{01}) when ${\bf{S}}_{m}\cdot{\bf{S}}_{n}\to S_m^zS_n^z$. Then, all terms in the Hamiltonian commute and, from such a perspective, one faces a classical system. Consider the Ising model on a bipartite lattice (such as square, simple-cubic etc.), which consists of two sublattices and all sites of one sublattice are surrounded by the sites of the other sublattice and vice versa. A rotation of the axes in the spin space for one sublattice results in the change of the  Ising interaction sign. Therefore, the thermodynamics (rather than correlations) does not depend on the sign of $J$. For an order-disorder phase transition this means that the Curie temperature equals the N\'{e}el temperature, i.e., $T_{\rm C}=T_{\rm N}$ as for the square-lattice Ising model $T_{\rm C}=T_{\rm N}\approx 0.567$ \cite{Onsager1944,Strecka2015} or for the honeycomb-lattice Ising model $T_{\rm C}=T_{\rm N}\approx 0.380$ \cite{Strecka2015}.
Frustrated Ising lattices are different. For example, the triangular-lattice ferromagnetic Ising model has the Curie temperature $T_{\rm C}\approx 0.910$ \cite{Strecka2015}, whereas its antiferromagnetic counterpart shows no order even at $T=0$ but, instead, has an exponentially large ground-state entropy \cite{Wannier1950,Wannier1973}. 
 
Recently, some interesting relations between the Heisenberg and Ising models on a frustrated lattice have been suggested in references~\cite{Popp2025,Ramirez2025}. The authors of these papers argued that the specific heat $c(T)$ of the frustrated-lattice Heisenberg model such as kagome, hyperkagome, etc., illustrates a hidden energy scale (which is considerably lower than $J$) as an extra low-temperature peak. Moreover, the accumulated entropy equals the residual ground-state entropy of the Ising model on the same lattice. 
 
\begin{figure}[h]
	\centering\includegraphics[width=0.35\columnwidth]{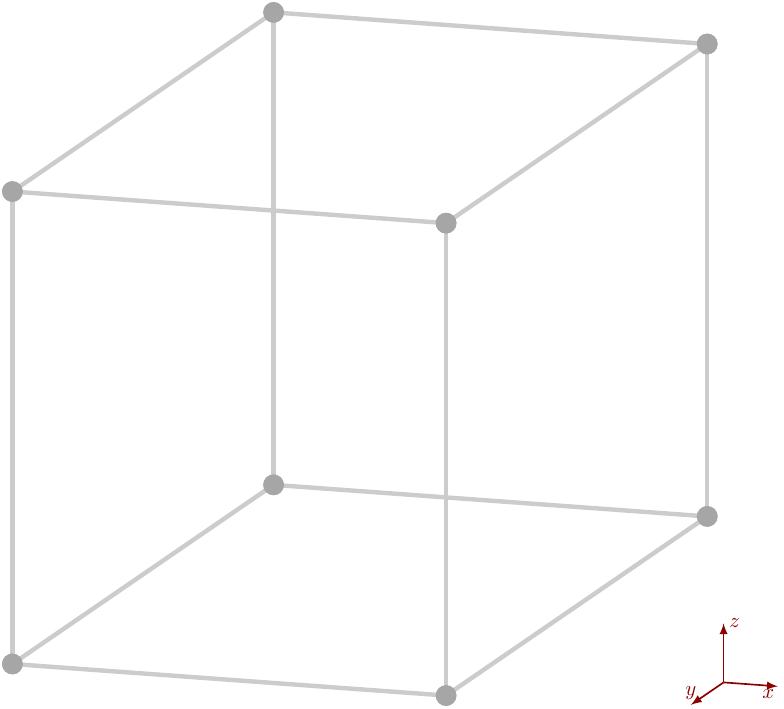}
	\centering\includegraphics[width=0.35\columnwidth]{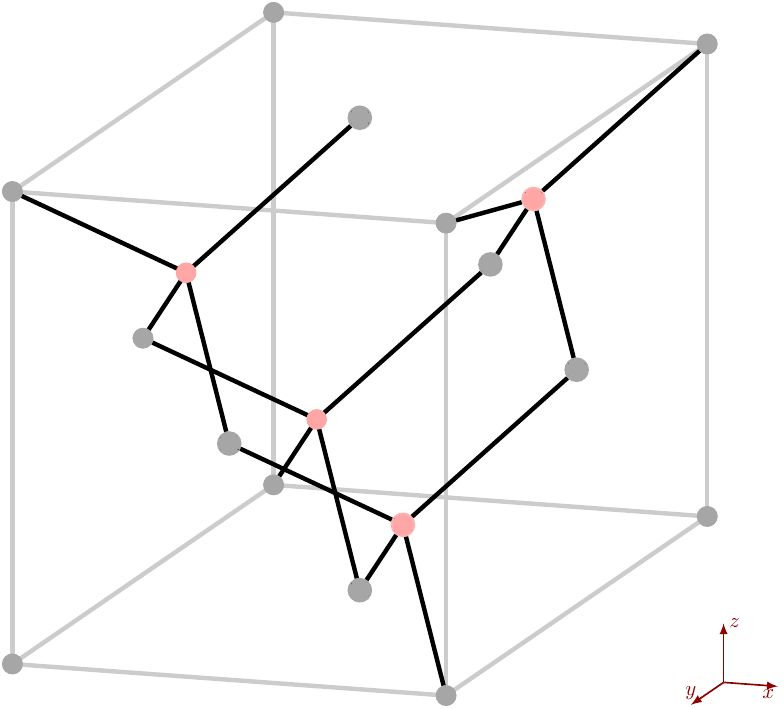}\\
	\vspace{3mm}
	\centering\includegraphics[width=0.35\columnwidth]{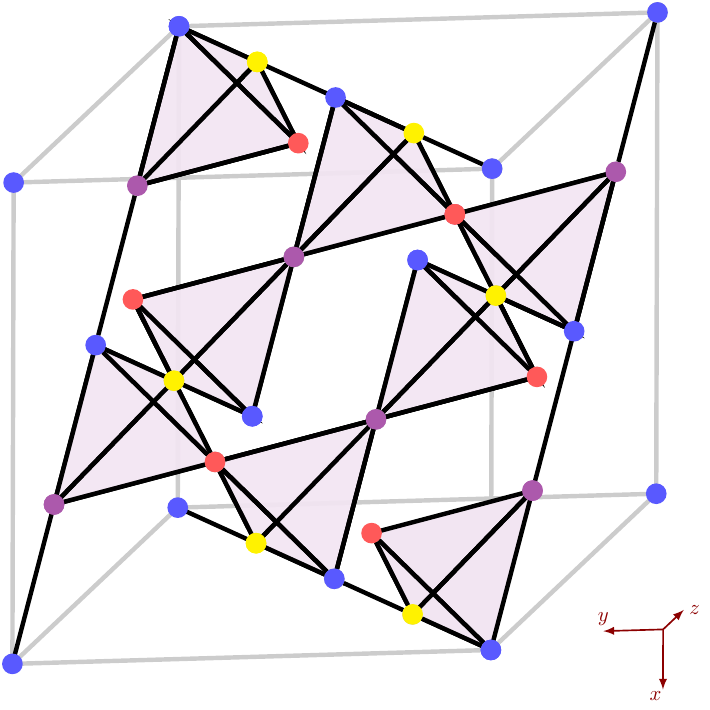}
	\centering\includegraphics[width=0.35\columnwidth]{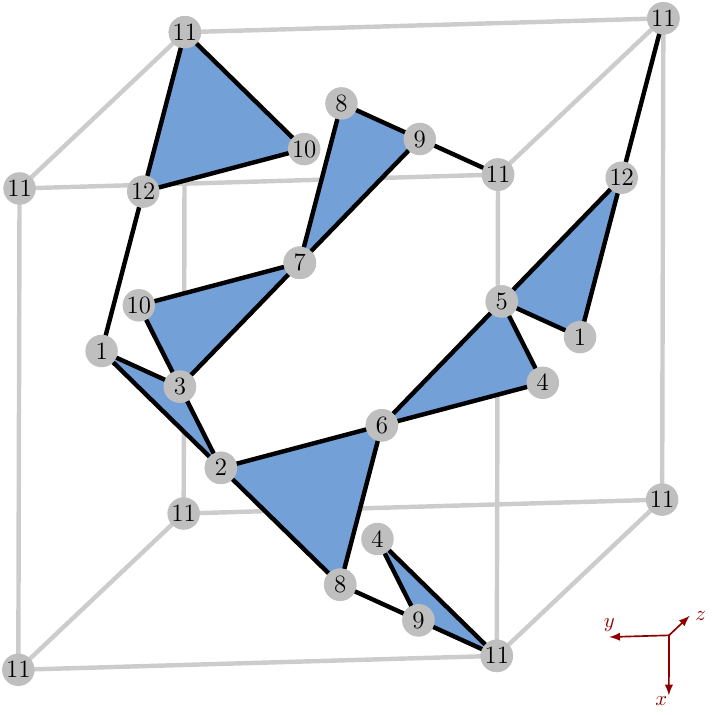}
	\caption{(Colour online) Simple-cubic (upper left), diamond (upper right), pyrochlore (lower left), and hyper\-kagome (lower right) lattices which are considered in this paper. Different colors of discs denote different sublattices, i.e., two sublattices for the diamond lattice and four sublattices for the pyrochlore lattice. The unit cell contains 1 (simple cubic), 2 (diamond), 4 (pyrochlore), or 12 (hyperkagome) sites (magnetic atoms). For a detailed description see the main text, section~\ref{s2a}.}
	\label{fig01}
\end{figure}

In this paper, we investigate the $S=1/2$ Heisenberg model on four three-dimensional lattices, see figure~\ref{fig01}. Two of them, simple-cubic and diamond, are bipartite ones, whereas other two, pyrochlore and hyperkagome, are geometrically frustrated ones. All these quantum spin-lattice systems are widely used in the theory of magnetism. Our goal is to discuss thermodynamic quantities for these models focusing on the specific heat in order to highlight the effect of lattice geometry in the features of $c(T)$ for the  quantum Heisenberg ferro- and antiferromagnets. To this end, we use quantum Monte Carlo simulations (ferromagnets and bipartite-lattice antiferromagnets) and high-temperature expansion series analysis (frustrated-lattice antiferromagnets). We have to notice that some of the presented results are not completely new but are scattered over literature, and here we have collected them in one place accompanied by still not reported results and all of these are considered from a common perspective.

The rest of the paper is organized as follows. 
In section~\ref{s2}, we introduce the models and methods.
Then, in section~\ref{s3}, we report our findings focusing on the temperature profiles of the specific heat.
Finally, we discuss and summarize in section~\ref{s4}.
Some additional information on the entropy-method calculations for the pyrochlore and hyperkagome cases is presented in appendix~\ref{App_A} and~\ref{App_B}, respectively. 

\section{Setting up the problem}
\label{s2}

\subsection{Lattices}
\label{s2a}

Let us describe the lattices to be considered, see figure~\ref{fig01}. Such descriptions are required for quantum Monte Carlo simulations using the ALPS package \cite{ALBUQUERQUE20071187,Bauer2011} as well as for generating high-temperature expansion series \cite{Pierre2024}.

The sites on the simple cubic lattice are defined by ${\bf{R}}_{\bf{n}}=n_{x}{\bf{e}}_{x}+n_{y}{\bf{e}}_{y}+n_{z}{\bf{e}}_{z}$, where ${\bf{e}}_{x}=(1,0,0)$, ${\bf{e}}_{y}=(0,1,0)$, ${\bf{e}}_{z}=(0,0,1)$ and $n_{x}$, $n_{y}$, $n_{z}$ are integers. Here and below, we set the length of the (unit-cell) cube edge to unity.

The diamond lattice consists of two interpenetrating face-centered cubic sublattices displaced by a vector $(1/4, 1/4, 1/4)$. That is, the sites on the diamond lattice are defined by ${\bf{R}}_{{\bf{m}}\alpha}={\bf{R}}_{\bf{m}}+{\bf{r}}_{\alpha}$. The sites of the face-centered-cubic lattice are determined by ${\bf{R}}_{\bf{m}}= m_1{\bf e}_1 + m_2{\bf e}_2 + m_3{\bf e}_3$, where ${\bf e}_1 = (0, 1/2, 1/2)$, ${\bf e}_2 = (1/2, 0, 1/2)$, ${\bf e}_3 = (1/2, 1/2, 0)$ and $m_1$, $m_2$, $m_3$ are integers; the origins of the two face-centered-cubic sublattices are taken to be ${\bf{r}}_1 = (0, 0, 0)$ and ${\bf{r}}_2 = (1/4, 1/4, 1/4)$.

The pyrochlore lattice consists of four interpenetrating face-centered-cubic sublattices with the lattice sites defined by ${\bf{R}}_{{\bf{m}}\alpha}={\bf{R}}_{\bf{m}}+{\bf{r}}_{\alpha}$. Here, the sites of the face-centered-cubic lattice are defined above, whereas the origins of the four face-centered-cubic sublattices are taken to be ${\bf{r}}_1 = (0, 0, 0)$, ${\bf{r}}_2 = (0, 1/4, 1/4)$, ${\bf{r}}_3 = (1/4, 0, 1/4)$, and ${\bf{r}}_4 = (1/4, 1/4, 0)$.

Finally, the hyperkagome lattice has 12 sites in a cubic unit cell. The sites on the hyperkagome lattice are defined by ${\bf{R}}_{\bf{n}\alpha}={\bf{R}}_{\bf{n}}+{\bf{r}}_{\alpha}$. Here, ${\bf{R}}_{\bf{n}}=n_{x}{\bf{e}}_{x}+n_{y}{\bf{e}}_{y}+n_{z}{\bf{e}}_{z}$ generates a simple-cubic-lattice translation (see above) and the origins of twelve equivalent sublattices are defined by ${\bf{r}}_{\alpha}$, $\alpha=1,2,\dots,12$, where
${\bf{r}}_{1}=(-1/2, 0, 1/2)$,
${\bf{r}}_{2}=(-1/4, 3/4, 1/2)$,
${\bf{r}}_{3}=(-1/2, 3/4, 1/4)$,
${\bf{r}}_{4}=(-1/4, 1/4, 0)$,
${\bf{r}}_{5}=(-1/2, 1/4, 3/4)$,
${\bf{r}}_{6}=(-1/4, 1/2, 3/4)$,
${\bf{r}}_{7}=(-3/4, 1/2, 1/4)$,
${\bf{r}}_{8}=( 0, 1/2, 1/2)$,
${\bf{r}}_{9}=( 0, 1/4, 1/4)$,
${\bf{r}}_{10}=(-3/4, 3/4, 0)$,
${\bf{r}}_{11}=( 0, 0, 0)$,
and
${\bf{r}}_{12}=(-3/4, 0, 3/4)$.

In quantum Monte Carlo simulations, we consider finite-size lattices with periodic boundary conditions, and the same linear extent ${\cal L}$ is taken in each lattice direction (a cube of periodicity), resulting in ${\cal N} = {\cal L}^3$ unit cells, or $N={\cal N}$ sites for the simple-cubic lattice, $N=2{\cal N}$ sites for the diamond lattice, $N=4{\cal N}$ sites for the pyroclore lattice, and $N=12{\cal N}$ sites for the hyperkagome lattice. While generating high-temperature expansion series, one defines the Heisenberg exchange interactions inside the unit cell and between the neighboring cells. In all cases the lattice geometry dictates the complexity of calculations and the required computation time.

\subsection{Methods}
\label{s2b}

\subsubsection{Quantum Monte Carlo}
\label{s2b1}

The $S=1/2$ ferromagnetic Heisenberg model can be studied using quantum Monte Carlo method. Moreover, the $S=1/2$ antiferromagnetic Heisenberg model on a bipartite lattice (such as simple-cubic or diamond) can be studied using quantum Monte Carlo method, too. We use the ALPS package, directed-loop scheme in stochastic series expansion method, to obtain temperature dependencies of the internal energy $E$, and calculate the specific heat by differentiation: $C(T)=\partial E/\partial T$. We consider finite systems of various sizes up to ${\cal L}=30$ or ${\cal L}=40$ with periodic boundary conditions imposed and switch on a small symmetry breaking field $h=10^{-4}$. Before averaging over 100\,000 steps, we perform 100\,000 steps for the initial relaxation to realize the equilibrium state.

The Monte Carlo approach is essentially a kind of stochastic simulation, the results of which depend on ${\cal L}$ (should be large), $h$ (should tend to zero), the number of thermalization and averaging steps (should be large); we tested various choices of these parameters to achieve reasonable outcomes at alowable computational cost. Finally, Monte Carlo simulations are usually accompanied by a finite-size scaling analysis, this way examining a critical behavior \cite{Mueller-Krumbhaar1986,Binder1989}, although the critical properties are not in the focus of the present study.

\subsubsection{High-temperature expansion series augmented by the entropy method}
\label{s2b2}

Using reference~\cite{Pierre2024}, we have generated the high-temperature expansion series for the specific heat (per site) $c(\beta)$, $\beta=1/T$,
\begin{eqnarray}
\label{06}
c(\beta)=\sum_{i=2}^{n}d_j\beta^i.
\end{eqnarray}
Here, for the pyrochlore lattice $n=17$, the coefficients $d_{2},\ldots,d_{13}$ were reported in reference~\cite{Derzhko2020} and $d_{14},\ldots,d_{17}$ in reference~\cite{Gonzalez2023}; for the hyperkagome lattice $n=20$ and the coefficients $d_{17},\ldots,d_{20}$,
\begin{eqnarray}
\label{07}
d_{17}=\frac{86\,356\,579\,620\,383}{7\,141\,645\,615\,104\,000},
\nonumber\\
d_{18}=\frac{79\,549\,971\,465\,578\,161}{1\,462\,609\,021\,973\,299\,200},
\nonumber\\
d_{19}=-\frac{1\,326\,067\,806\,520\,547\,891}{124\,321\,766\,867\,730\,432\,000},
\nonumber\\
d_{20}=-\frac{680\,190\,389\,551\,589\,559\,239}{17\,902\,334\,428\,953\,182\,208\,000},
\end{eqnarray}
complement those that were reported in reference~\cite{Singh2012}, see also appendix~\ref{App_B}. 

Within the entropy method \cite{Bernu2001,Misguich2005,Bernu2015}, one works with the entropy as a function of the energy $s(e)$, $e_0\leqslant e\leqslant 0$, $e_0$ is the ground-state energy, and the Maclaurin series for $s(e)$ is known from equation~(\ref{06}). Assuming gapples low-energy excitations (they yield a power-law decay of the specific heat $c(T)$ as $T\to 0$ with the exponent $\alpha$), one interpolates an auxiliary function $G(e)=[s(e)]^{(1+\alpha)/\alpha}/(e-e_0)$ between $e=e_0$ and $e=0$ using Pad\'{e} approximants $[u, d](e)$ with $d+u\leqslant n$, $n=17$ (pyrochlore) or $n=20$ (hyperkagome): $G(e)\approx[(\ln 2)^{\alpha/(1+\alpha)}/(-e_0)]\,[u, d](e)$. As a result, one gets the approximate entropy $s(e)\approx [(e-e_0)G(e)]^{\alpha/(1+\alpha)}$ and therefore, the specific heat $c(T)$ in the parametric form: $T=1/s^{\prime}(e)$, $c(e)=-[s^{\prime}(e)]^2/s^{\prime\prime}(e)$; here $(\ldots)^{\prime}={\rm d}(\ldots)/{\rm d}e$.
Thus, besides high-temperature expansion series (\ref{06}), one needs the values of $e_0$ and $\alpha$. Under another assumption, when low-energy excitations are gapped, one works with another auxiliary function $G(e)=(e-e_0)[s(e)/(e-e_0)]^{\prime}\approx (\ln 2/e_0)[u,d](e)$, which yields the approximate entropy $s(e)/(e-e_0)\approx-\ln 2/e_0-\int_e^0{\rm d}\xi G(\xi)/(\xi-e_0)$, and one needs as an input the value of $e_0$ only. 

It should be noted that even if the required input information is unavailable, one may still proceed with the entropy method and find those parameters that are missing searching for such their values which yield the largest number of close to each other $c(T)$ based on various Pad\'{e} approximants $[u,d](e)$ \cite{Bernu2020}. Such a heuristic criteria were used in 2020 \cite{Derzhko2020} in the study of the pyrochlore-lattice $S=1/2$ Heisenberg antiferromagnet on the basis of 13th order high-temperature expansion series to arrive at $e_0\approx -0.52$ and $\alpha=2$ as a most favorable set of the unknown parameters. For the hyperkagome-lattice case, similar arguments lead to $e_0$ which lies between $-0.441$ and $-0.435$ and $\alpha=2$ \cite{Hutak2024}. 

Since 2021, there have been several proposals for the ground-state energy of the pyrochlore-lattice $S=1/2$ Heisenberg antiferromagnet: 
$-0.490(6)$ (density-matrix renormalization group method \cite{Hagymasi2021}), 
$-0.477(3)$ (variational Monte Carlo methods \cite{Astrakhantsev2021}). 
Besides, more terms in high-temperature expansion series have become available now \cite{Gonzalez2023}.
Furthermore, for the hyperkagome-lattice $S=1/2$ Heisenberg antiferromagnet, four more coefficients in equation~(\ref{06}) are available now in comparison to reference~\cite{Singh2012}, see equation~(\ref{07}).
All this motivates us to reconsider the entropy-method calculations for the pyrochlore- and hyperkagome-lattice $S=1/2$ Heisenberg antiferromagnets and refine the temperature profile $c(T)$, see section~\ref{s3b2} below.

\section{Specific heat $c(T)$}
\label{s3}

\subsection{Ferromagnets}
\label{s3a}

\begin{figure}[h]
	\centering\includegraphics[scale=0.67]{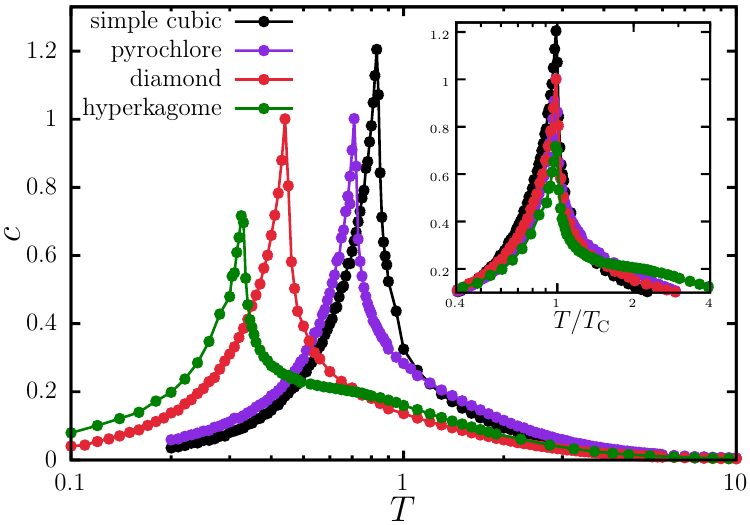} 
	\caption{(Colour online) Specific heat $c(T)$ for the $S=1/2$ Heisenberg ferromagnet on four three-dimensional lattices. Quantum Monte Carlo simulations for the number of unit cells ${\cal L}=40$ (simple cubic) and ${\cal L}=30$ (pyrochlore, diamond, hyperkagome) and a small symmetry breaking field $h=10^{-4}$. Inset presents $c$ as a function of $T/T_{\rm C}$.}
	\label{fig02}
\end{figure}

\begin{figure}[h!]
	\centering\includegraphics[scale=0.67]{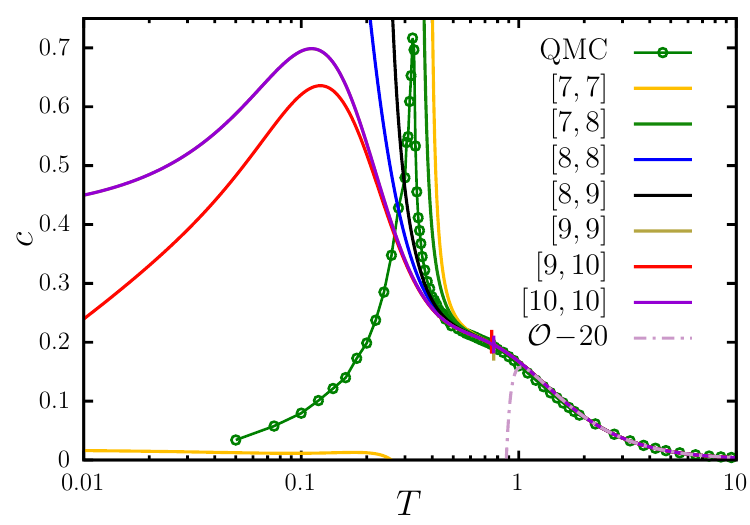} 
	\caption{(Colour online) Towards a sloped shoulder in the paramagnetic phase for the hyperkagome-lattice $S = 1/2$ Heisenberg ferromagnet: high-order Pad\'{e} approximants $[u,d](T)$, $u + d \leqslant 20$, to the specific heat series $c(\beta)$ versus quantum Monte Carlo simulations (which were reported in figure~\ref{fig02}).}
	\label{fig02new03}
\end{figure}

Our findings for ferromagnets are collected in the middle column of table~\ref{tab1} and in figure~\ref{fig02}.
As is generally known, spin waves or magnons are low-lying excitations and therefore  $c(T)\propto T^{3/2}$ for all ferromagnets. An order-disorder phase transition occurs at the Curie temperature $T_{\rm C}$, see table~\ref{tab1}. Note that $T_{\rm C}$ strongly depends on the lattice geometry. Around $T_{\rm C}$, one expects the critical behavior, $c(T)-c(T_{\rm C})\propto\vert T-T_{\rm C}\vert^{-\alpha}$, $\alpha\approx -0.12$ (3D Heisenberg universality class, see reference~\cite{Kivelson2024}). Here, the critical exponent $\alpha<0$ and the specific heat  is finite at $T_{\rm C}$. In paramagnetic phase, as the temperature tends to infinity, all quantum Heisenberg feromagnets become the Curie paramagnet. Note a peculiar behavior for the hyperkagome-lattice case: $c(T)$ exhibits a sloped shoulder in the paramagnetic phase (green curves in figure~\ref{fig02}). The existence of the sloped shoulder in $c(T)$ is also confirmed by simple Padé approximants to the high-temperature series for $c(\beta)$ (\ref{06}), compare $[10,10](T)$, $[9,10](T)$, and $[9,9](T)$ [indistinguishable from $[10,10](T)$] to quantum Monte Carlo data in figure~\ref{fig02new03}. Meanwhile we do not have a rigorous and convincing explanation for that and this issue needs further study. 

\begin{table}[h]
	\caption{\label{tab1} Curie point $T_{\rm C}$ and N\'{e}el point $T_{\rm N}$ for several three-dimensional lattices considered in this paper.}
	\vspace{1em}
	\centering
	\begin{tabular}{lll}
		& $T_{\rm C}$ ($J=-1$)                  & $T_{\rm N}$ ($J=1$)           \\
		\hline
		simple-cubic &$0.839(1)$ \cite{Wessel2010}           &$0.946(1)$ \cite{Wessel2010}   \\
		\hline
		diamond      &$0.447\pm 0.001$ \cite{Oitmaa2018}     &$0.531\pm 0.001$ \cite{Oitmaa2018} \\
		&$0.445$ \cite{Kuzmin2019}              &                               \\
		&$0.444\,47(5)$ \cite{Barwolf2025}      &$0.527\,82(5)$ \cite{Barwolf2025} \\
		\hline
		pyrochlore	 &$0.718$ \cite{Muller2017}              & --                            \\
		\hline
		hyperkagome  &$0.33$ \cite{Parymuda2024,Parymuda2025}& --                            \\
	\end{tabular}
\end{table}

\subsection{Antiferromagnets}
\label{s3b}

\subsubsection{Bipartite lattices}
\label{s3b1}

\begin{figure}[h]
	\centering\includegraphics[scale=0.67]{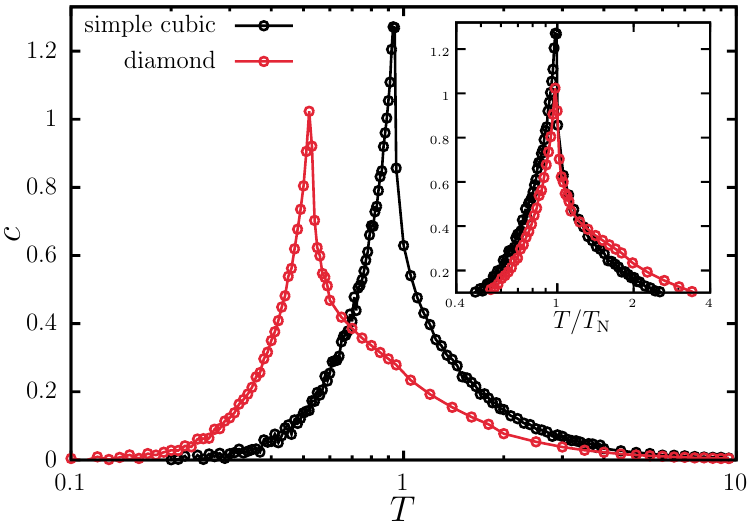} 
	\caption{(Colour online) Specific heat $c(T)$ for the $S=1/2$ Heisenberg antiferromagnet on two bipartite three-dimensional lattices. Quantum Monte Carlo simulations for the number of unit cells ${\cal L}=40$ (simple cubic) and ${\cal L}=30$ (diamond) and a small symmetry breaking field $h=10^{-4}$. Inset presents $c$ as a function of $T/T_{\rm N}$.}
	\label{fig03}
\end{figure}

Our findings for bipartite-lattice antiferromagnets are collected in the right-hand column of table~\ref{tab1} and in figure~\ref{fig03}. Again, low-lying excitations are spin waves with linear dispersion relation, leading to a power-law decay of the specific heat as $T\to 0$ with the exponent $3$. With temperature increase, the antiferromagnetic order is gradually suppressed by thermal fluctuations and is completely destroyed at the N\'{e}el temperature $T_{\rm N}$ (see table~\ref{tab1}) and above it. Note that $T_{\rm C}<T_{\rm N}$ as expected for the quantum $S = 1/2$ Heisenberg magnets on bipartite lattices, see, e.g., reference~\cite{Oitmaa2004}.

\subsubsection{Frustrated lattices}
\label{s3b2}

\begin{figure}[h]
	\centering\includegraphics[scale=0.65]{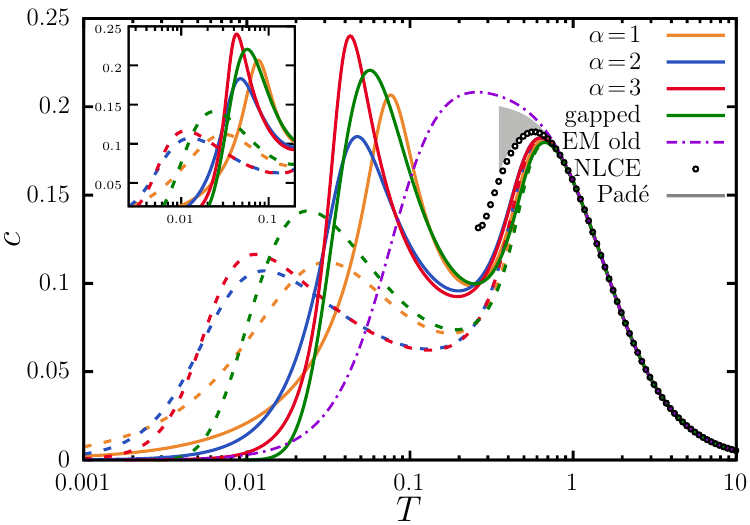} 
	\centering\includegraphics[scale=0.65]{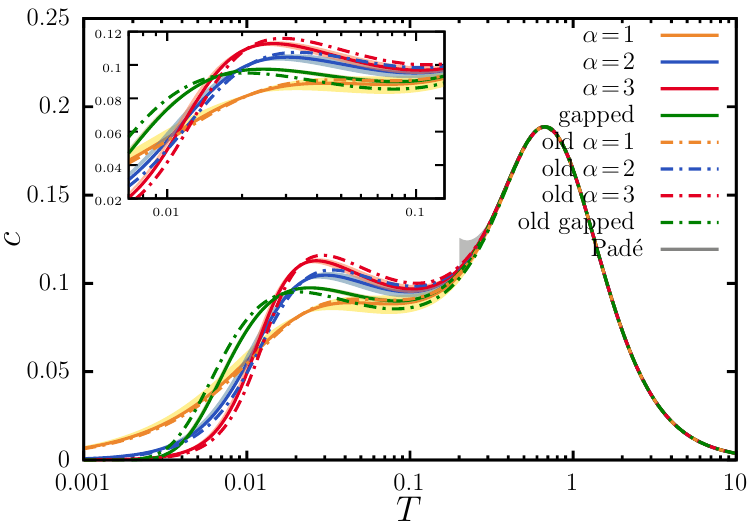} 
	\caption{(Colour online) Specific heat $c(T)$ for the $S=1/2$ Heisenberg antiferromagnet on two frustrated  three-dimensional lattices. Entropy method results. Top: pyrochlore lattice. Solid curves correspond to $e_0=-0.49$ \cite{Hagymasi2021} whereas dashed curves correspond to $e_0=-0.477$ \cite{Astrakhantsev2021} with $[8,9](e)$ approximant; dash-dotted curve is taken from reference~\cite{Derzhko2020}. In addition, we present numerical-linked-cluster-expansion data taken from \cite{Schaefer2020} (circles) as well as simple Pad\'{e} approximants $[8,9](T)$ and $[8,8](T)$ which constitute the boundary of the gray area. Bottom: hyperkagome lattice. The ground-state energy is determined within the entropy method and we took $[10,10](e)$ approximant to obtain all these entropy-method curves; dash-dotted curves are taken from the previous study \cite{Hutak2024}. In addition, we present simple Pad\'{e} approximants $[8,9](T)$ and $[10,10](T)$ which constitute the boundary of the gray area.}
	\label{fig04}
\end{figure}

Frustrated-lattice antiferromagnets are less understood. As said above, there is no universal method to get reliable results for $c(T)$: quantum Monte Carlo fails because of infamous sign problem, exact diagonalization or finite-temperature Lanczos methods are restricted to too small lattices in three spatial dimensions, the rotation-invariant Green’s function method, although is principally applicable, contains uncontrolled approximation. High-temperature expansion series approach may be useful, but only if its applicability is extended to sufficiently low temperatures. Therefore, we have performed the entropy-method calculations, see section~\ref{s2b2}. Our findings are reported in figure~\ref{fig04}. 

First of all we note that we have extended previous calculations \cite{Derzhko2020,Hutak2024} by (i) including more terms (17 instead of 13 (pyrochlore) or 20 instead of 16 (hyperkagome)) and (ii) using the present value of the ground-state energy $e_0$ for the pyrochlore-lattice case. For the hyperkagome-lattice $S=1/2$ Heisenberg antiferromagnet we obtain within the entropy method a more precise estimate for the ground-state energy $e_0\in[-0.4396,-0.4358]$ (according to reference~\cite{Hutak2024} $e_0\in[-0.441,-0.435]$).

For geometrically frustrated Heisenberg magnets, no magnetic order occurs until zero temperature. (Note, however, that some symmetries for the pyrochlore-lattice case are broken in the ground state, see references~\cite{Hagymasi2021,Astrakhantsev2021,Hering2022}.) From figure~\ref{fig04}, the lack of a clear magnetic ordering feature in $c(T)$ is evident: $c(T)$ shows a much broader peak (bump or rounded maxima), than the usually seen peak (sharp cusp) for a magnetic transition, at $T$ somewhat below $J=1$, compare figure~\ref{fig04} to figure~\ref{fig03}. (Note also that the scale for $c$ is now about 5 times smaller.) Besides the main maximum of the order of $J$, there is a low-temperature maximum, to be discussed below.

We have to comment what is new in figure~\ref{fig04} in comparison to references~\cite{Derzhko2020,Hutak2024}. Concerning the pyrochlore case, in the upper panel of figure~\ref{fig04} we show the ``old'' entropy-method curve $c(T)$ obtained in reference~\cite{Derzhko2020} [dash-dotted, $n=13$, $e_0=-0.52$, $\alpha=2$, $[6,6](e)$], the result of the numerical linked cluster expansion \cite{Schaefer2020} (empty circles, above $T\approx 0.25$), as well as two Pad\'{e} approximants $[8,9](T)$ and $[8,8](T)$ (gray shaded area, above $T=0.35$). ``New'' entropy-method curves are based on $n=17$ terms in equation~(\ref{06}), $e_0=-0.49$ (\cite{Hagymasi2021}, solid) or $e_0=-0.477$ (\cite{Astrakhantsev2021}, dashed), $[8,9](e)$ approximant, and different assumptions about a law for vanishing $c(T)$ as $T\to 0$, see also appendix~\ref{App_A}. Obvious ambiguity in the shape of $c(T)$ is related to the lack of knowledge about the ground state and low-lying excitations.
Concerning the hyperkagome case, in the lower panel of figure~\ref{fig04} we show, besides the entropy-method curves, two Pad\'{e} approximants $[8,9](T)$ and $[10,10](T)$ (gray shaded area, above $T=0.2$). The entropy-method curves which follow from $[10,10](e)$ approximant for an assumed scenario of low-temperature behavior of $c(T)$ are accompanied by shadow areas. The boundaries of this area correspond to two (slightly) different values of $e_0$ which give the same (large) number of almost coinciding curves $c(T)$ obtained from Pad\'{e} approximants $[u,d](e)$, for details see reference~\cite{Hutak2024} and also appendix~\ref{App_B}. The new curves in the lower panel of figure~\ref{fig04} (solid) are quite close to those reported in reference~\cite{Hutak2024} (dash-dotted).
\newpage

\subsubsection{Low-temperature peak and a hidden energy scale}
\label{s3b3}

Recently, a theory of a low-temperature peak of the specific heat for frustrated quantum Heisenberg antiferromagnets was suggested in references~\cite{Popp2025,Ramirez2025}.
These authors considered various geometrically frustrated lattices such as kagome, triangular, hyperkagome, or spinel to emphasize the existence of two temperature scales which manifest themselves in two distinct, well-distinguishable peaks in $c(T)$ and argued that the low-temperature peak comes from nonmagnetic excitations, similar to spin exchanges in the chains of spins, whose energy is well separated from the energy of other magnetic excitations, similar to spin flips, which give rise to the high-temperature peak in $c(T)$. Both panels of figure~\ref{fig04} with the  entropy-method findings for $c(T)$ qualitatively agree  with such a picture. Moreover, the authors of references~\cite{Ramirez2025,Popp2025} also predicted the entropy of the excitations contributing to the low-temperature peak: since the excitations that give rise to that peak are adiabatically connected to the ground states of the Ising model on the same lattice, the associated entropy $s(T^*)$ should match the entropy of the ground states of that Ising model~${s}(0)$. 
Here, the entropy related to the low-temperature peak may be estimated as $s(T^*)=\int_{0}^{T^*}{\rm d}Tc(T)/T$, i.e., assuming that the low-temperature peak spreads up to $T^*$, whereas the residual ground-state entropy of the Ising model ${s}(0)=\lim_{T\to 0}{s}(T)$, ${s}(T)=\int_{0}^{T}{\rm d}T{c}(T)/T=\ln 2-\int_{T}^{\infty}{\rm d}T{c}(T)/T$.

To check this theory, we have to compare the low-temperature peak entropy for the Heisenberg model~$s(T^*)$ with the residual ground-state entropy for the Ising model ${s}(0)$. The latter quantity, ${s}(0)$, follows from rigorous analytical calculations in the two-dimensional case, e.g., references~\cite{Wannier1950,Wannier1973,Kano1953} or from approximate analytical and numerical calculations in the three-dimensional case, e.g., references~\cite{Anderson1956,Yoshioka2004}. The former quantity, $s(T^*)$, is related to $c(T)$, which, in turn, is known only from numerics and mainly in the two-dimensional case, e.g., references~\cite{Schnack2018,Bernu2020}. (In reference~\cite{Popp2025} the authors numerically compute~$c(T)$ of a kagome cluster of $N=18$ sites.)

With the obtained temperature profiles shown in figure~\ref{fig04} we may further discuss the theory of the hidden energy scale of reference~\cite{Popp2025} using as a testing bed two three-dimensional systems. According to figure~\ref{fig04}, for the entropy under the lower temperature peak we get:
\begin{eqnarray}
\label{08}
s(T^*)\in [0.325,0.350]
\end{eqnarray}
(pyrochlore, $T^*=0.2$, $\alpha=1,2,3$ and gapped, $e_0=-0.49$ \cite{Hagymasi2021} and $e_0=-0.477$ \cite{Astrakhantsev2021})
and
\begin{eqnarray}
\label{09}
s(T^*)\in [0.244,0.257] 
\end{eqnarray}
(hyperkagome, $T^*=0.1$, $\alpha=1,2,3$ and gapped).

\begin{figure}[h]
	\centering\includegraphics[scale=0.66]{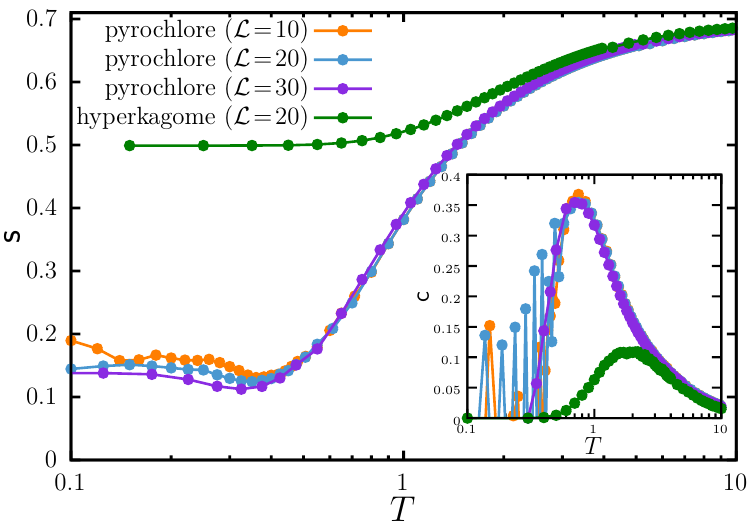}
	\caption{(Colour online) Towards the residual ground-state entropy of the uniaxial Ising model on the pyrochlore and hyperkagome lattices. The entropy is obtained by integration from the Monte Carlo data for the specific heat (up to ${\cal L}=30$), shown in the inset. Again, $h=10^{-4}$.}
	\label{fig05}
\end{figure}

On the other hand, we need the residual ground-state entropy of the Ising antiferromagnet on the pyrochlore and hyperkagome lattices. To this end, we consider the $S=1/2$ models with the uniaxial Ising-like anisotropy and perform classical Monte Carlo simulations \cite{ALBUQUERQUE20071187,Bauer2011} for the specific heat ${C}(T)=\partial {E}/\partial T$ to obtain the entropy ${s}(T)$ per site through the formula ${s}(T)=\ln 2-\int_T^{\infty}{\rm d}T{c}(T)/T$. Our findings are reported in figure~\ref{fig05}. We obtain ${s}(0)\approx 0.1$ (pyrochlore) and ${s}(0)\approx 0.5$ (hyperkagome). 
We may also mention here reference~\cite{Anderson1956} (see also \cite{Bramwell1998}), where some calculations using Pauling's technique to estimate the zero-point entropy were reported. 
Moreover, similar classical Monte Carlo simulations were presented in reference~\cite{Yoshioka2004}, although for  the $S=1/2$ models with the Ising-like (i.e., easy-axis) anisotropy along the direction into the center of the tetrahedron (pyrochlore) or along the line joining the triangular centers (hyperkagome). The result for the hyperkagome case just very slightly exceeds the one in figure~\ref{fig05}, but the result for the pyrochlore case is noticeably larger than the one in figure~\ref{fig05} and the residual ground-state entropies differ approximately by a factor of about 2.
It is worth noting that from the physical point of view the Ising-like anisotropy (spin space) is expected to agree with the crystal symmetry (real space). The uniaxial anisotropy requires a unique crystalline axis, which is not available in the three-dimensional pyrochlore and hyperkagome lattices. By contrast, the other  above mentioned Ising-like anisotropies are consistent with the lattice symmetries. For more discussion on global-axis and local-axis Ising models see reference~\cite{Pohle2023}.

Comparing both the data for the pyrochlore-lattice case, $0.325\ldots0.350$ versus $0.1/0.2$, and the hyperkagome-lattice case, $0.244\ldots0.257$ versus $0.5$, we observe that these numbers are not very close. Of course, we have to note an ambiguity in estimation of $s(T^*)$ in equations~(\ref{08}) and (\ref{09}) simply by taking $T^{*}=0.2$ and $T^{*}=0.1$. Besides, the residual ground-state entropy of the Ising models obtained by classical Monte Carlo simulations needs a better accuracy. One may also ask which exactly two-state Ising model, uniaxial or the one consistent with crystalline axis, should be examined: Both are acceptable for adiabatic connection with the Heisenberg model but have different residual ground-state entropies. Probably the most important comment comes after taking a closer look at the hidden energy scale theory of references~\cite{Ramirez2025,Popp2025}. The authors considered, as a starting point, the Ising model on the lattice at hand which has the degeneracy of ground states that scales exponentially with the lattice size. Then, switching on the transverse-coupling Hamiltonian, which is controlled by the anisotropy parameter $\delta$, $0\leqslant \delta\leqslant 1$, lifts the ground-state degeneracy and results in a set of low-lying excitations which is well separated from the spin-flip-like excitations until $\delta$ is close to 0. These low-lying (i.e., low-energy) excitations are responsible for a low-temperature peak of $c(T)$, which is well separated from the main high-temperature peak of $c(T)$. And the entropy associated with the low-temperature peak of $c(T)$ should match the residual ground-state entropy of the Ising model on the lattice under consideration. However, it is not clear a priori whether the low-lying excitations constitute a set of states  well separated from other excitations if $\delta\to 1$. To end up, a better understanding of the hidden energy scale, in particular, at the quantitative level awaits further research.

\section{Discussion and summary}
\label{s4}

In this study, we presented the specific heat probes of several quantum three-dimensional Heisenberg magnets. Our main findings are reported in figures~\ref{fig02}, \ref{fig03}, and \ref{fig04}. While ferromagnets and bipartite-lattice antiferromagnets can be straightforwardly investigated by applying quantum Monte Carlo simulations, frustrated-lattice antiferromagnets require more sophisticate techniques: We use high-temperature expansion series complemented by the entropy-method interpolation. Our findings illustrate the effects of geometry on the ordering temperature (ferromagnets and bipartite-lattice antiferromagnets). For frustrated-lattice antiferromagnets, the obtained specific heat, which interpolates between the assumed low-energy properties and rigorously known high-temperature ones, provides some information about a possible energy spectrum of the quantum spin system at hand.  
Although our study is pure theoretical, it may be useful from the solid-state perspective, if the corresponding compounds are available. 

\section*{Data availability statement}

The data that support the findings of this study are available from the authors upon reasonable request.

\section*{Acknowledgements}

The authors are thankful to the Armed Forces of Ukraine for protection since 2014, and especially since February 24, 2022. This project is funded by the National Research Foundation of Ukraine (2023.03/0063, Frustrated quantum magnets under various external conditions). The authors are much indebted to Laura Messio for carrying out a test calculation for the hyperkagome lattice using the Paris high-temperature-series-expansions code \cite{Pierre2024}. 
O.~D. thanks the Abdus Salam International Centre for Theoretical Physics (Trieste) for kind hospitality at the Joint ICTP-WE Heraeus School and Workshop on Advances in Quantum Matter: Pushing the Boundaries, 4--15 August 2025.

\appendix
\section{Pyrochlore-lattice antiferromagnet} 
\label{App_A}
\renewcommand{\theequation}{A.\arabic{equation}}
\setcounter{equation}{0}

\renewcommand{\thefigure}{A.\arabic{figure}}
\setcounter{figure}{0}

\begin{figure}[h]
	\centering\includegraphics[scale=0.67]{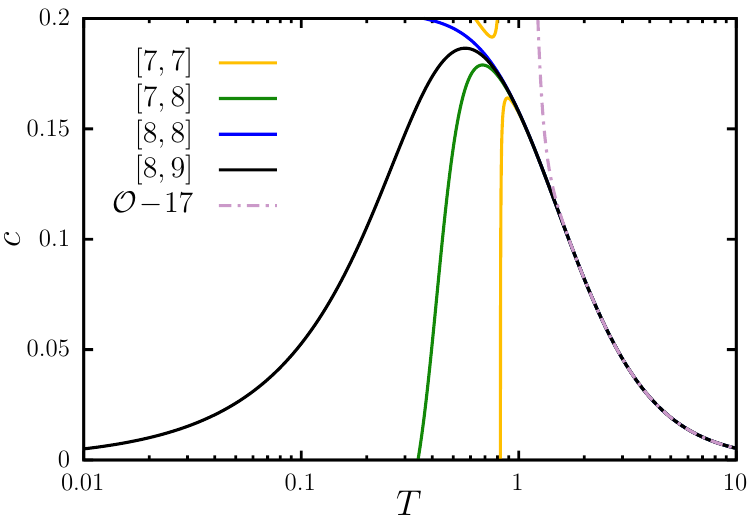}
	\caption{(Colour online) Pad\'{e} approximants to high-temperature expansion series for the $S=1/2$ Heisenberg antiferromagnet on pyrochlore lattice.}
	\label{fig06}
\end{figure}

For the pyrochlore-lattice antiferromagnet, the coefficients $d_2,\ldots,d_{17}$ are reported in reference~\cite{Gonzalez2023} (the first thirteen coefficients can be also found in reference~\cite{Derzhko2020}). Simple Pad\'{e} approximants to high-temperature expansion series (\ref{06}) are shown in figure~\ref{fig06}. As can be seen from this figure, they can provide reliable results above $T\approx0.7$.

To apply the entropy-method interpolation, we use the ground-state energy $e_0$ found in referen\-ces~\cite{Hagymasi2021,Astrakhantsev2021} and make an assumption about the decay of the specific heat as $T\to 0$. We try a power-law decay with several exponents $\alpha=1,2,3$, as well as an exponential decay. However, none of the made assumptions yields an excellent temperature profile $c(T)$ with respect to large number $n_{\rm cP}$ of almost coinciding $c(T)$ which follow from $n_{\rm P}=u+d-3=14$ different $[u,d](e)$ approximants. On the other hand, we may try to determine $e_0$ as in reference~\cite{Bernu2020}. However, we get $n_{\rm cP}=7$ for $e_0=-0.423\,4,\ldots,-0.420\,2$ if $\alpha=1$, $n_{\rm cP}=6$ for $e_0=-0.461\,3,\ldots,-0.42$ if $\alpha=2$, or $n_{\rm cP}=7$ for $e_0=-0.465\,2,\ldots,-0.457\,8$ if $\alpha=3$; these values are quite far from $e_0=-0.49$ \cite{Hagymasi2021} and $e_0=-0.477$ \cite{Astrakhantsev2021}. Apparently, the calculation of $c(T)$ for the pyrochlore-lattice $S=1/2$ Heisenberg antiferromagnet is far from being finally settled.

\section{Hyperkagome-lattice antiferromagnet} 
\label{App_B}
\renewcommand{\theequation}{B.\arabic{equation}}
\setcounter{equation}{0}

\renewcommand{\thefigure}{B.\arabic{figure}}
\setcounter{figure}{0}

We begin with the coefficients $d_2,\ldots,d_{16}$ in equation~(\ref{06}), which are as follows:
\begin{eqnarray}
\label{b1}
d_{2}&=&\frac{3}{8},
\quad
d_{3}=0,
\quad
d_{4}=-\frac{51}{128},
\quad
d_{5}=0,
\quad
d_{6}=\frac{349}{1\,024},
\nonumber\\
d_{7}&=&-\frac{7}{1\,280},
\quad
d_{8}=-\frac{131\,449}{491\,520},
\nonumber\\
d_{9}&=&\frac{349}{35\,840},
\quad
d_{10}=\frac{11\,103\,797}{55\,050\,240},
\nonumber\\
d_{11}&=&-\frac{6\,080\,063}{495\,452\,160},
\quad
d_{12}=-\frac{981\,828\,121}{6\,606\,028\,800},
\nonumber\\
d_{13}&=&\frac{192\,742\,927}{14\,533\,263\,360},
\quad
d_{14}=\frac{2\,254\,101\,727\,553}{20\,927\,899\,238\,400},
\nonumber\\
d_{15}&=&-\frac{142\,269\,385\,877}{10\,882\,507\,603\,968},
\nonumber\\
d_{16}&=&-\frac{2\,346\,255\,743\,077\,553}{30\,471\,021\,291\,110\,400}.
\end{eqnarray}
These coefficients can be found, in principle, in reference~\cite{Singh2012}, where they are presented, although as decimal fractions. In equation~(\ref{b1}) they are given as common fractions which may be sometime more convenient.

\begin{figure}[h]
	\centering\includegraphics[scale=0.65]{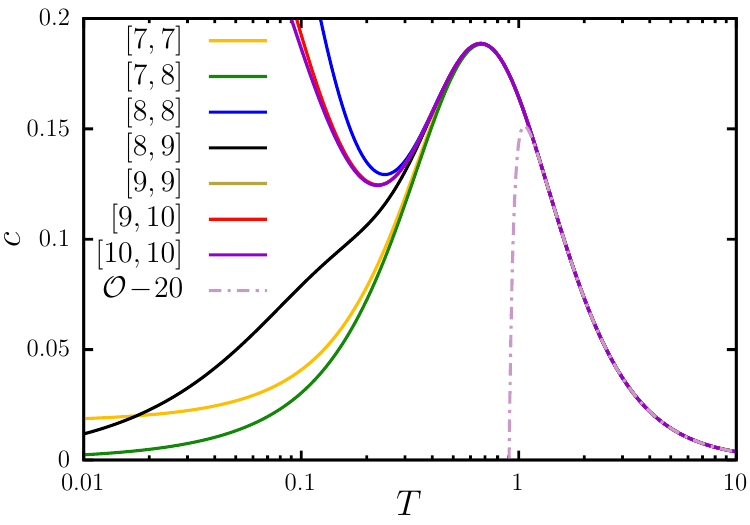}
	\centering\includegraphics[scale=0.65]{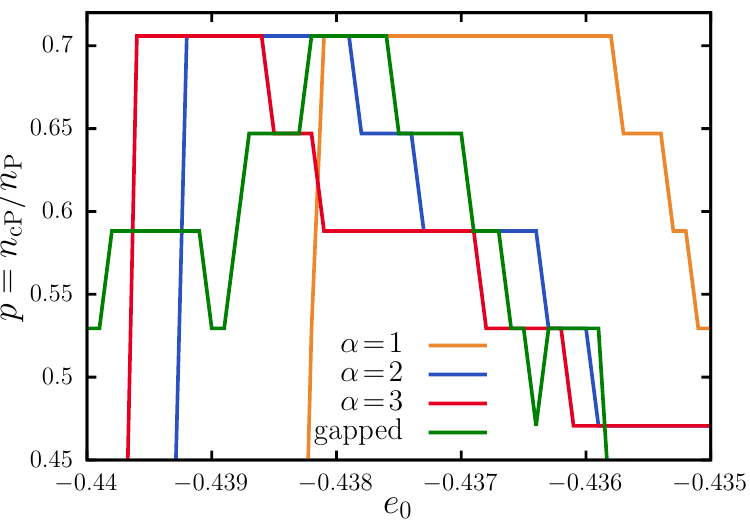}
	\caption{(Colour online) Entropy-method calculations for the $S=1/2$ Heisenberg antiferromagnet on hyperkagome lattice. Top: Pad\'{e} approximants of high-temperature expansion series. Bottom: ground-state energy determination within the entropy method.}
	\label{fig07}
\end{figure}

The entropy method calculations are described in detail in reference~\cite{Hutak2024}, but now we use as an input 20 coefficients in equation~(\ref{06}) (instead of 16 in reference~\cite{Hutak2024}). From the upper panel of figure~\ref{fig07} one concludes that Pad\'{e} approximants of high-temperature expansion series can give $c(T)$ for $T$ exceeding $\approx 0.3$. In the lower panel of figure~\ref{fig07} we closely follow reference~\cite{Hutak2024} and present the number of almost coinciding $c(T)$ denoted as $n_{\rm cP}$ among all considered $n_{\rm P}=u+d-3$ Pad\'{e} approximants $[u,d](e)$ as a function of the chosen ground-state energy $e_0$ (this quantity is unknown) for various scenarios. Next, we use the range determined this way for $e_0$ and calculate $c(T)$ using $[10,10](e)$ approximant to obtain the curves and the shaded areas in the lower panel of figure~\ref{fig04}. For example, for $\alpha=2$, the values of $e_0$ determining the blue curve and shaded area in the lower panel of figure~\ref{fig04} are as follows: $e_0=-0.4392$, $e_0=-0.4386$, and $e_0=-0.4379$, cf. the lower panel of figure~\ref{fig07}.

\bibliographystyle{cmpj}
\bibliography{thermod_refsOL}

\begin{thebibliography}{10}
\providecommand{\url}[1]{\texttt{#1}}
\providecommand{\urlprefix}{URL }
\expandafter\ifx\csname urlstyle\endcsname\relax
  \providecommand{\doi}[1]{doi:\discretionary{}{}{}#1}\else
  \providecommand{\doi}{doi:\discretionary{}{}{}\begingroup
  \urlstyle{rm}\Url}\fi
\providecommand{\eprint}[2][]{\url{#2}}

\bibitem{Born2013}
Born~M., {Atomic Physics}, Dover Publications, 2013.

\bibitem{Vakarchuk2012}
Vakarchuk~I.~O., {Quantum Mechanics}, Ivan Franko National University of Lviv,
  Lviv, 2012 (in Ukrainian).

\bibitem{Duncan2019}
Duncan~A., Janssen~M., {Constructing Quantum Mechanics},
  Oxford~University~Press, 2019.

\bibitem{Onsager1944}
Onsager~L., Phys. Rev., 1944, \textbf{65}, 117--149,
  \doi{10.1103/PhysRev.65.117}.

\bibitem{Strecka2015}
Stre\ifmmode~\check{c}\else \v{c}\fi{}ka~J., Ja\ifmmode \check{s}\else
  \v{s}\fi{}\ifmmode~\check{c}\else \v{c}\fi{}ur~M., Acta Phys. Slovaca, 2015,
  \textbf{65}, 235--367.

\bibitem{Wannier1950}
Wannier~G.~H., Phys. Rev., 1950, \textbf{79}, 357--364,
  \doi{10.1103/PhysRev.79.357}.

\bibitem{Wannier1973}
Wannier~G.~H., Phys. Rev. B, 1973, \textbf{7}, 5017--5017,
  \doi{10.1103/PhysRevB.7.5017}.

\bibitem{Popp2025}
Popp~P., Ramirez~A.~P., Syzranov~S., Phys. Rev. Lett., 2025, \textbf{134},
  226701, \doi{10.1103/PhysRevLett.134.226701}.

\bibitem{Ramirez2025}
Ramirez~A.~P., Syzranov~S.~V., Mater. Adv., 2025, \textbf{6}, 1213--1229,
  \doi{10.1039/D4MA00914B}.

\bibitem{ALBUQUERQUE20071187}
Albuquerque~A., Alet~F., Corboz~P., Dayal~P., Feiguin~A., Fuchs~S., Gamper~L.,
  Gull~E., G{\"u}rtler~S., Honecker~A., et~al., J. Magn. Magn. Mater., 2007,
  \textbf{310}, No.~2, 1187--1193, \doi{10.1016/j.jmmm.2006.10.304}.

\bibitem{Bauer2011}
Bauer~B., Carr~L.~D., Evertz~H.~G., Feiguin~A., Freire~J., Fuchs~S., Gamper~L.,
  Gukelberger~J., Gull~E., Guertler~S., et~al., J. Stat. Mech.: Theory Exp.,
  2011, \textbf{2011}, No.~05, P05001, \doi{10.1088/1742-5468/2011/05/P05001}.

\bibitem{Pierre2024}
Pierre~L., Bernu~B., Messio~L., SciPost Phys., 2024, \textbf{17}, 105,
  \doi{10.21468/SciPostPhys.17.4.105}.

\bibitem{Mueller-Krumbhaar1986}
M{\"u}ller-Krumbhaar~H., Simulation of Small Systems, Springer Berlin
  Heidelberg, Berlin, Heidelberg, 1986, 195--223,
  \doi{10.1007/978-3-642-82803-4_5}.

\bibitem{Binder1989}
Binder~K., Wang~J.~S., J. Stat. Phys., 1989, \textbf{55}, 87,
  \doi{10.1007/BF01042592}.

\bibitem{Derzhko2020}
Derzhko~O., Hutak~T., Krokhmalskii~T., Schnack~J., Richter~J., Phys. Rev. B,
  2020, \textbf{101}, 174426, \doi{10.1103/PhysRevB.101.174426}.

\bibitem{Gonzalez2023}
Gonzalez~M.~G., Bernu~B., Pierre~L., Messio~L., Phys. Rev. B, 2023,
  \textbf{107}, 235151, \doi{10.1103/PhysRevB.107.235151}.

\bibitem{Singh2012}
Singh~R. R.~P., Oitmaa~J., Phys. Rev. B, 2012, \textbf{85}, 104406,
  \doi{10.1103/PhysRevB.85.104406}.

\bibitem{Bernu2001}
Bernu~B., Misguich~G., Phys. Rev. B, 2001, \textbf{63}, 134409,
  \doi{10.1103/PhysRevB.63.134409}.

\bibitem{Misguich2005}
Misguich~G., Bernu~B., Phys. Rev. B, 2005, \textbf{71}, 014417,
  \doi{10.1103/PhysRevB.71.014417}.

\bibitem{Bernu2015}
Bernu~B., Lhuillier~C., Phys. Rev. Lett., 2015, \textbf{114}, 057201,
  \doi{10.1103/PhysRevLett.114.057201}.

\bibitem{Bernu2020}
Bernu~B., Pierre~L., Essafi~K., Messio~L., Phys. Rev. B, 2020, \textbf{101},
  140403, \doi{10.1103/PhysRevB.101.140403}.

\bibitem{Hutak2024}
Hutak~T., Krokhmalskii~T., Schnack~J., Richter~J., Derzhko~O., Phys. Rev. B,
  2024, \textbf{110}, 054428, \doi{10.1103/PhysRevB.110.054428}.

\bibitem{Hagymasi2021}
Hagym\'asi~I., Sch\"afer~R., Moessner~R., Luitz~D.~J., Phys. Rev. Lett., 2021,
  \textbf{126}, 117204, \doi{10.1103/PhysRevLett.126.117204}.

\bibitem{Astrakhantsev2021}
Astrakhantsev~N., Westerhout~T., Tiwari~A., Choo~K., Chen~A., Fischer~M.~H.,
  Carleo~G., Neupert~T., Phys.~Rev.~X, 2021, \textbf{11}, 041021,
  \doi{10.1103/PhysRevX.11.041021}.

\bibitem{Kivelson2024}
Kivelson~S.~A., Jiang~J.~M., Chang~J., {Statistical Mechanics of Phases and
  Phase Transitions}, Princeton~University~Press, 2024.

\bibitem{Wessel2010}
Wessel~S., Phys. Rev. B, 2010, \textbf{81}, 052405,
  \doi{10.1103/PhysRevB.81.052405}.

\bibitem{Oitmaa2018}
Oitmaa~J., J. Phys.: Condens. Matter, 2018, \textbf{30}, No.~15, 155801,
  \doi{10.1088/1361-648X/aab22c}.

\bibitem{Kuzmin2019}
Kuz'min~M.~D., Philos. Mag. Lett., 2019, \textbf{99}, No.~9, 338--350,
  \doi{10.1080/09500839.2019.1692156}.

\bibitem{Barwolf2025}
B\"arwolf~R., Sushchyev~A., Parisen~Toldin~F., Wessel~S., Phys. Rev. B, 2025,
  \textbf{111}, 085136, \doi{10.1103/PhysRevB.111.085136}.

\bibitem{Muller2017}
M\"uller~P., Lohmann~A., Richter~J., Menchyshyn~O., Derzhko~O., Phys. Rev. B,
  2017, \textbf{96}, 174419, \doi{10.1103/PhysRevB.96.174419}.

\bibitem{Parymuda2024}
Parymuda~M., Krokhmalskii~T., Derzhko~O., Preprint of the Institute for
  Condensed Matter Physics, ICMP--24--03E, Lviv, 2024.

\bibitem{Parymuda2025}
Parymuda~M., Krokhmalskii~T., Derzhko~O., J. Phys.: Condens. Matter, 2025,
  \textbf{37}, No.~33, 335801, \doi{10.1088/1361-648X/adf67d}.

\bibitem{Oitmaa2004}
Oitmaa~J., Zheng~W., J. Phys.: Condens. Matter, 2004, \textbf{16}, No.~47,
  8653, \doi{10.1088/0953-8984/16/47/016}.

\bibitem{Hering2022}
Hering~M., Noculak~V., Ferrari~F., Iqbal~Y., Reuther~J., Phys. Rev. B, 2022,
  \textbf{105}, 054426, \doi{10.1103/PhysRevB.105.054426}.

\bibitem{Schaefer2020}
Sch\"afer~R., Hagym\'asi~I., Moessner~R., Luitz~D.~J., Phys. Rev. B, 2020,
  \textbf{102}, 054408, \doi{10.1103/PhysRevB.102.054408}.

\bibitem{Kano1953}
Kan\^o~K., Naya~S., Prog. Theor. Phys., 1953, \textbf{10}, No.~2, 158--172,
  \doi{10.1143/ptp/10.2.158}.

\bibitem{Anderson1956}
Anderson~P.~W., Phys. Rev., 1956, \textbf{102}, 1008--1013,
  \doi{10.1103/PhysRev.102.1008}.

\bibitem{Yoshioka2004}
Yoshioka~T., Koga~A., Kawakami~N., J. Phys. Soc. Jpn., 2004, \textbf{73},
  No.~7, 1805--1811, \doi{10.1143/JPSJ.73.1805}.

\bibitem{Schnack2018}
Schnack~J., Schulenburg~J., Richter~J., Phys. Rev. B, 2018, \textbf{98},
  094423, \doi{10.1103/PhysRevB.98.094423}.

\bibitem{Bramwell1998}
Bramwell~S.~T., Harris~M.~J., J. Phys.: Condens. Matter, 1998, \textbf{10},
  No.~14, L215, \doi{10.1088/0953-8984/10/14/002}.

\bibitem{Pohle2023}
Pohle~R., Jaubert~L. D.~C., Phys. Rev. B, 2023, \textbf{108}, 024411,
  \doi{10.1103/PhysRevB.108.024411}.

\end{thebibliography}

\ukrainianpart

\title{Тривимірні нефрустровані та фрустровані квантові Гайзенбергові магнетики. Дослідження теплоємності}
\author{Т. Крохмальский\refaddr{label1}, Т. Гутак\refaddr{label1}, О. Держко\refaddr{label1,label2}}
\addresses{
	\addr{label1} Iнститут фiзики конденсованих систем імені І. Р. Юхновського Нацiональної академiї наук України,
	79011, м. Львiв, вул.~Свєнцiцького, 1, Україна
	\addr{label2} Кафедра теоретичної фізики імені професора Iвана Вакарчука, Львівський національний університет імені Івана Франка, 
	79005, м. Львiв, вул.~Драгоманова, 12, Україна
}

\makeukrtitle

\begin{abstract}
	\tolerance=3000%
	Ми досліджуємо $S=1/2$ магнетик Гайзенберга на чотирьох тривимірних гратках --- простій кубічній, алмазу, пірохлору і гіперкагоме --- для феромагнітного і антиферомагнітного знаків обмінної взаємодії, щоб проілюструвати ефект геометрії гратки на скiнченнотемпературнi термодинамічні властивості, зокрема, на теплоємність $c(T)$. З цією метою ми використовуємо симуляції методом квантового Монте Карло і високотемпературні розвинення, доповненні методом ентропії. Ми також обговорюємо недавню пропозицію про прихований енергетичний масштаб у геометрично фрустрованих магнетиках.

	\keywords квантова Гайзенбергова спінова модель, геометрично фрустровані гратки
\end{abstract}
\lastpage
\end{document}